\documentclass[11pt]{article}
\usepackage{bbold}
\usepackage{geometry}
\usepackage{amsmath}

\usepackage{amsfonts}
\usepackage{amssymb}
\usepackage{graphicx}
\usepackage{caption}
\usepackage{subcaption}
\usepackage{hyperref}
\usepackage{float}
\usepackage{braket}
\usepackage{mathrsfs}
\usepackage{fancyhdr}
\title{\vspace{-1cm}\textbf{Time-dependent metric for the two dimensional, non-Hermitian coupled oscillator}}
\date{}
\author{\textbf{Andreas Fring and Thomas Frith}\\\textit{Department of Mathematics, City, University of London,}\\
	\textit{Northampton Square, London EC1V 0HB, UK}\\
	\textit{E-mail: a.fring@city.ac.uk, thomas.frith@city.ac.uk}}
\numberwithin{equation}{section}

\begin{document}

\maketitle 
\thispagestyle{fancy}
ABSTRACT: We provide a time-dependent Dyson map and metric for the two dimensional harmonic oscillator with a non-Hermitian $i xy$ coupling term. This particular time-independent model exhibits spontaneously broken $\mathcal{PT}$-symmetry and becomes unphysical in the broken regime, with the spectrum becoming partially complex. By introducing an explicit time-dependence into the Dyson map, we provide a time-dependent metric that renders the model consistent across the unbroken and broken regimes.
\section{Introduction}
The central defining quantities in quantum mechanics are the Hamiltonian, $\mathcal{H}$ and the quantum wavefunction, $\varphi$. These objects are important for describing the behaviour of any quantum system, but they are only fully defined when associated with a Hilbert space. This statement is true for both Hermitian and non-Hermitian Hamiltonians. In addition, the associated Hilbert space requires a well defined metric, which in the Hermitian case is usually chosen in a very suggestive way, but is nonetheless selected. In the non-Hermitian case the choice is less obvious and even allows for the possibility to choose a time-dependent metric \cite{ExactSols,fring2016non,fring2016unitary}. 

When this non-Hermitian Hamiltonian is time-dependent, then a time-dependent metric seems quite suggestive, but is also possible when the non-Hermitian Hamiltonian is time-independent. This allows for the interesting possibility of starting with a time-independent non-Hermitian Hamiltonian and relating it to a time-dependent Hermitian Hamiltonian via a well defined, time-dependent metric. Furthermore, this construction can even hold when the non-Hermitian Hamiltonian is in its broken regime \cite{FRING20172318,HigherDims}. This regime is usually discarded as the energy eigenvalues become complex \cite{bender1998real,bender2007making,mostafazadeh2010pseudo,moiseyev2011non}

In this paper we demonstrate how a time-independent, non-Hermitian Hamiltonian with spontaneously broken $\mathcal{PT}$-symmetry can be consistently associated to a Hilbert space across all regimes by using a time-dependent metric. We begin by defining two Hamiltonian systems, one being non-Hermitian  and time-independent $H^\dagger\neq H$, and the other being Hermitian and time-dependent, $h\left(t\right)^\dagger=h\left(t\right)$ together with their associated wavefunctions. We therefore have both Hamiltonians satisfying their own time-dependent Schr\"odinger equation (TDSE)
\begin{equation}
i\hbar\partial_t\psi\left(t\right)=H\psi\left(t\right), \quad i\hbar\partial_t\phi\left(t\right)=h\left(t\right)\phi\left(t\right).
\end{equation}
Next we make the proposition that the wavefunctions are related by the time-dependent Dyson map $\eta\left(t\right)$
\begin{equation}
\phi\left(t\right)=\eta\left(t\right)\psi\left(t\right).
\end{equation}
Substituting the expression for $\phi\left(t\right)$ into the Schr\"odinger equation, we arrive at the time-dependent Dyson equation (TDDE) \cite{maamache2017pseudo,znojil2008time,mostafazadeh2007time,de2006time,mostafazadeh2018energy}
\begin{equation}\label{TDDE}
h\left(t\right)=\eta\left(t\right)H\eta^{-1}\left(t\right)+i\hbar\partial_t\eta\left(t\right)\eta^{-1}\eta\left(t\right).
\end{equation} 
We construct the metric from the Dyson map $\varrho\left(t\right)=\eta\left(t\right)^\dagger\eta\left(t\right)$. The calculation of $\varrho\left(t\right)$ allows one to root the non-Hermitian theory in a Hilbert space, meaning one can define the inner product as well calculate observables \cite{bender1998real,bender2007making,mostafazadeh2010pseudo}. In what follows we set $\hbar=1$.
\newpage
\section{2D non-Hermitian coupled harmonic oscillator}
\rhead{}
\lhead{Time-dependent metric for the 2D, non-Hermitian coupled oscillator}
In this paper we analyse a popular time-independent, two dimensional, non-Hermitan, $\mathcal{PT}$-symmetric Hamiltonian that exhibits spontaneous symmetry breaking.
\begin{equation}\label{Hamiltonian}
H=\frac{1}{2m}\left(p_x^2+p_y^2\right)+\frac{1}{2}m\left(\Omega_x^2x^2+\Omega_y^2y^2\right)+i\lambda xy \qquad\qquad \lambda, \Omega_x, \Omega_y, m \in \mathbb{R},
\end{equation}
This Hamiltonian has energy eigenvalues
\begin{equation}
E_{n_1,n_2}=\left(n_1+\frac{1}{2}\right)\omega_x+\left(n_2+\frac{1}{2}\right)\omega_y,
\end{equation} 
with
\begin{equation}
\omega_{x,y}^2=\frac{1}{2m}\left(m\Omega_+^2\pm\sqrt{m^2\Omega_-^4-4\lambda^2}\right),
\end{equation}
and $\Omega_\pm^2=\Omega_y^2\pm\Omega_x^2$. This system has been studied in detail in the time-independent regime \cite{HigherDims,PhysRevA.91.062101,MANDAL20131043} and it is clear from the energy eigenvalues that the $\mathcal{PT}$-symmetry is spontaneously broken when $|m\Omega_-^2|<2|\lambda|$ as they become complex. We show in this paper that by using a time-dependent metric, we can mend the broken regime and return a physical meaning to this regime.

We wish to express the Hamiltonian (\ref{Hamiltonian}) in terms of a closed algebra. This allows us to formulate our Ansatz for $\eta$ in terms of the generators of the algebra and guarantees that the resulting Hermitian Hamiltonian will also be expressible in terms of these generators. The algebra for our Hamiltonian is comprised of the ten Hermitian generators
\begin{equation}
\begin{aligned}
\begin{split}
K_{\pm}^{z}=&\frac{1}{2}\left(p_z^2\pm z^2\right),\quad K_0^z=\frac{1}{2}\{z,p_z\},\quad J_\pm=\frac{1}{2}\left(xp_y\pm yp_x\right), \quad I_\pm=\frac{1}{2}\left(xy\pm p_xp_y\right),\\
\end{split}
\end{aligned}
\end{equation}
where $z=x,y$. The commutation relations for these generators are
\begin{eqnarray}
\left[K_0^z,K_\pm^z\right]\!\!\!\!&=&\!\!\!\!2iK_\mp^z,\quad \left[K_+^z,K_-^z\right]=2iK_0^z,\quad \left[K_{\mu}^x,K_{\nu}^y\right]=0,\\
\left[K_0^x,J_\pm\right]\!\!\!\!&=&\!\!\!\!-iJ_\mp, \,\,\,\,\quad\left[K_0^y,J_\pm\right]=iJ_\mp, \qquad\,\, \left[K_0^z,I_\pm\right] =- iI_\mp,  \\
\left[K_\pm^x,J_{+}\right]\!\!\!\!&=&\!\!\!\!\pm iI_{\mp}, \!\!\!\qquad\qquad\qquad\qquad\qquad\quad \left[K_\pm^y,J_{+}\right]=\pm iI_{\mp}\\
\left[K_\pm^x,J_{-}\right]\!\!\!\!&=&\!\!\!\!\mp iI_{\pm}, \!\!\!\qquad\qquad\qquad\qquad\qquad\quad \left[K_\pm^y,J_{-}\right]=\pm iI_{\pm}\\
\left[K_\pm^x,I_{+}\right]\!\!\!\!&=&\!\!\!\!\pm iJ_{\mp}, \!\!\!\qquad\qquad\qquad\qquad\qquad\quad \left[K_\pm^y,I_{+}\right]=-iJ_{\mp},\\ 
\left[K_\pm^x,I_{-}\right]\!\!\!\!&=&\!\!\!\!\mp iJ_{\pm}, \!\!\!\qquad\qquad\qquad\qquad\qquad\quad \left[K_\pm^y,I_{-}\right]=-iJ_{\pm},\\ 
\left[J_+,J_-\right]\!\!\!\!&=&\!\!\!\!\frac{i}{2}\left(K_0^x-K_0^y\right), \!\!\qquad\qquad\qquad\qquad \left[I_+,I_-\right]=-\frac{i}{2}\left(K_0^x+K_0^y\right),\\
\left[J_{+},I_\pm\right]\!\!\!\!&=&\!\!\!\!\pm\frac{i}{2}\left(K_{\mp}^x+K_{\mp}^y\right), \!\!\!\qquad\qquad\qquad\quad
\left[J_{-},I_\pm\right]=\mp\frac{i}{2}\left(K_{\pm}^x-K_{\pm}^y\right),\\
\end{eqnarray}
with $\mu,\nu=+,-,0$. As is clear, this is a rich algebra containing many closed sub-algebras contained within. We can rewrite the Hamiltonian (\ref{Hamiltonian}) as
\begin{equation}
H=\sum_{z,\sigma=\pm}\Lambda^z_\sigma K^z_\sigma+i\lambda\left(I_++I_-\right),
\end{equation}
where $\Lambda^z_\pm=\frac{1}{2m}\left(1\pm m^2\Omega_z^2\right)$. As the generators are all Hermitian, the Hamiltonian is non-Hermitian due to the contribution from the last term. With our Hamiltonian expressed in this form we are now able to proceed with solving the time-dependent Dyson equation using the Baker-Campbell-Hausdorff formula to evaluate the adjoint action of $\eta$ on the Hamiltonian and solving the resulting differential equations. These equations arise when we enforce the condition of Hermiticity on the resulting Hamiltonian. However, we first recall the solution to time-independent Dyson equation \cite{HigherDims} in order to emphasise that the mapping and the metric breaks down as the $\mathcal{PT}$-symmetry is spontaneously broken in the absence of time.

\section{Time-independent Dyson map}
In the time-independent case, the TDDE (\ref{TDDE}) reduces to a similarity transformation and is solved with \cite{HigherDims}
\begin{equation}
\eta=e^{\theta J_-}, \quad \tanh2\theta=\frac{2\lambda}{m\Omega_-^2}.
\end{equation}
This mapping is only valid for $|m\Omega_-^2|>2|\lambda|$ which matches the results from \cite{PhysRevA.91.062101,MANDAL20131043}. The resulting Hermitian Hamiltonian is
\begin{equation}
h=\frac{1}{2m}\left(p_x^2+p_y^2\right)+\frac{1}{2}m\left(\omega_x^2x^2+\omega_y^2y^2\right),
\end{equation}
\begin{equation}
\omega_x^2=\frac{\Omega_x^2\cosh^2\theta+\Omega_y^2\sinh^2\theta}{\cosh2\theta}, \quad \omega_y^2=\frac{\Omega_x^2\sinh^2\theta+\Omega_y^2\cosh^2\theta}{\cosh2\theta}.
\end{equation}
In this time-independent setting, when the $\mathcal{PT}$-symmetry is broken we cannot construct a metric $\varrho=\eta^\dagger\eta$ and therefore cannot make sense of the broken regime. To progress, we must acknowledge that our choice for $\eta$, and therefore $\varrho$, is not restricted to be time-independent. Introducing an explicit time-dependence into these parameters means we are led to solve the TDDE resulting in a time-dependent Hermitian Hamiltonian.
\section{Time-dependent Dyson map}
We now use a time-dependent Dyson map of the form
\begin{equation}\label{dysonmap}
\eta\left(t\right)=e^{\alpha_-\left(t\right)L_-}e^{\theta_+\left(t\right) J_+}e^{\alpha_+\left(t\right)L_+}e^{\theta_-\left(t\right) J_-}, \quad L_+=\frac{1}{2}\left(I_++I_-\right),\quad L_-=\frac{1}{2}\left(I_+-I_-\right).
\end{equation}
The quantities $\alpha_+$, $\alpha_-$, $\theta_+$ and $\theta_-$ are real. This Ansatz is of course is not the most general choice. We could use all ten generators in our Ansatz and use complex coefficients. However, it is well known that $\eta$ is not unique \cite{scholtz1992quasi} and so we are content here to find \textbf{a} solution. We choose (\ref{dysonmap}) to be comprised of the interaction generators between the two dimensions as this produces a comprehensible solution. Substituting $\eta$ into the TDDE (\ref{TDDE}), the imaginary terms and eliminated when the following differential equations hold
\begin{equation}\label{diffeq}
\begin{aligned}
\begin{split}
\dot{\alpha}_-=&-\frac{2}{m}\sin\theta_+,\\
\dot{\theta}_+=&\frac{1}{4m}\alpha_-\left(2m^2\Omega_+^2-\alpha_+^2\right)\sec\theta_+-\frac{\alpha_+}{m},\\
\dot{\alpha}_+=&\frac{1}{2m}\left(2m^2\Omega_+^2-\alpha_+^2\right)\tan\theta_++m\Omega_-^2\sinh\theta_--2\lambda\cosh\theta_-,\\
\dot{\theta}_-=&\frac{\alpha_-\left(2\lambda\sinh\theta_--m\Omega_-^2\cosh\theta_-\right)}{2\cos\theta_++\alpha_+\alpha_-},
\end{split}
\end{aligned}
\end{equation}
with the overdot denoting the time-derivative. We solve these coupled differential equations by differentiating the first equation three times and at each stage substituting in the expressions for $\dot{\theta}_+$, $\dot{\theta}_-$, $\dot{\alpha}_+$ and $\dot{\alpha}_-$
\begin{equation}\label{alphameq}
\begin{aligned}
\begin{split}
\ddot{\alpha}_-&=\frac{1}{2m^2}\alpha_+\left(4\cos\theta_++\alpha_+\alpha_-\right)-\Omega_+^2\alpha_-,\\
\dddot{\alpha}_-&=\frac{1}{m^2}\left(m\Omega_-^2\sinh\theta_--2\lambda\cosh\theta_-\right)\left(2\cos\alpha_++\alpha_+\alpha_-\right)+\frac{4\Omega_+^2}{m}\sin\alpha_+,\\
\ddddot{\alpha}_-&=-2\Omega_+^2\left[\frac{1}{2m^2}\alpha_+\left(4\cos\theta_++\alpha_+\alpha_-\right)-\Omega_+^2\alpha_-\right]-\left(\Omega_-^4-4\frac{\lambda^2}{m^2}\right)\alpha_-.
\end{split}
\end{aligned}
\end{equation}
In the final, fourth order equation we can clearly substitute in $\ddot{\alpha}_-$ to obtain a fourth order equation solely in terms of $\alpha_-$
\begin{equation}\label{FourthOrder}
\ddddot{\alpha}_-+2\Omega_+^2\ddot{\alpha}_-+\delta\alpha_-=0,
\end{equation}
where $\delta=\Omega_-^4-4\frac{\lambda^2}{m^2}$. The solution to equation (\ref{FourthOrder}) allows us to go back and calculate $\theta_+$, $\theta_-$ and $\alpha_+$ using the equations (\ref{diffeq}) and (\ref{alphameq}) in terms of $\alpha_-$ together with its derivatives. 
\begin{equation}
\theta_+=-\arcsin \frac{m\dot{\alpha}_-}{2}, \quad \alpha_+=\frac{-\sqrt{4-m^2\dot{\alpha}^2_-}\pm\beta}{\alpha_-}, \quad \theta_-=\log\left[\frac{m^2\gamma\pm m\sqrt{m^2\gamma^2+\delta\beta^2}}{\beta\left(m\Omega_-^2-2\lambda\right)}\right],
\end{equation}
where $\beta=\sqrt{4+2m^2\Omega^2_+\alpha^2_--m^2\left(\dot{\alpha}^2_--2\alpha_-\ddot{\alpha}_-\right)}$ and $\gamma=2\Omega_+^2\dot{\alpha}_-+\dddot{\alpha}_-$.
Solving equation (\ref{FourthOrder}) therefore completes the solution for the Dyson map. 

We can solve (\ref{FourthOrder}) without consideration to the sign of $\delta$ and obtain a valid solution. However, we wish to preserve the reality of $\alpha_-$ in order to prevent $\eta$ from being a simple gauge transformation. Therefore we must consider three separate regimes arising from the time-independent analysis, these are: the unbroken regime where $|m\Omega_-^2|>2|\lambda|$ ($\delta>0$), the spontaneously broken regime with $|m\Omega_-^2|<2|\lambda|$ ($\delta<0$) and the exceptional point where $|m\Omega_-^2|=2|\lambda|$ ($\delta=0$). These regimes must be treated separately as they lead to qualitatively different solutions. In all three cases $\alpha_-\in\mathbb{R}$ as required.

For $\delta>0$, the solution is
\begin{equation}
\alpha_-=c_1\cos\left(\Delta_+ t\right)+c_2\sin\left(\Delta_+ t\right)+c_3\cos\left(\Delta_- t\right)+c_4\sin\left(\Delta_- t\right),
\end{equation}
where  $c_{1,2,3,4}$ are constants of integration. The number of constants reflects the number of first order differential equations we started with in (\ref{diffeq}), and so we get four as expected. The frequencies are
\begin{equation}
\Delta_\pm=\sqrt{\Omega_+^2\pm2\sqrt{\Omega_x^2\Omega_y^2+\frac{\lambda^2}{m^2}}}.
\end{equation}
The term in the inner square root is always postive. The condition for $\Delta_-$ to be real is $\delta>0$, so in the unbroken regime both $\Delta_\pm$ are real. When $\delta<0$, $\Delta_-$ becomes imaginary and so solving (\ref{FourthOrder}) in broken regime must be considered separately.

For $\delta<0$, the solution is
\begin{equation}
\alpha_-=\tilde{c}_1\cos\left(\tilde{\Delta}_+ t\right)+\tilde{c}_2\sin\left(\tilde{\Delta}_+ t\right)+\tilde{c}_3\cosh\left(\tilde{\Delta}_- t\right)+\tilde{c}_4\sinh\left(\tilde{\Delta}_- t\right)
\end{equation} 
with $\tilde{c}_{1,2,3,4}$ being the constants of integration and 
\begin{equation}
\tilde{\Delta}_\pm=\sqrt{2\sqrt{\Omega_x^2\Omega_y^2+\frac{\lambda^2}{m^2}}\pm\Omega_+^2}.
\end{equation} 
We have both $\tilde{\Delta}_\pm$ being real when $\delta<0$. Therefore even in the broken regime, we obtain a real solution for $\alpha_-$ and consequently for $\eta$. 

Finally, for $\delta=0$ the solution is
\begin{equation}
\alpha_-=\hat{c}_1\cos\left(\sqrt{2}\Omega_+ t\right)+\hat{c}_2\sin\left(\sqrt{2}\Omega_+ t\right)+\hat{c}_3t+\hat{c}_4,
\end{equation}
where $\hat{c}_{1,2,3,4}$ are the constants of integration.

We have obtained a real solution for $\alpha_-$ for all values of $\delta$. This means we have a real, well-defined metric, $\varrho\left(t\right)=\eta\left(t\right)^\dagger\eta\left(t\right)$, for all values of $\Omega_x$, $\Omega_y$, $\lambda$ and $m$
\begin{equation}
\varrho\left(t\right)=e^{\theta_-\left(t\right)J_-}e^{\alpha_+\left(t\right) L_+}e^{\theta_+\left(t\right)J_+}e^{2\alpha_-\left(t\right)L_-}e^{\theta_+\left(t\right) J_+}e^{\alpha_+\left(t\right)L_+}e^{\theta_-\left(t\right) J_-}.
\end{equation}
Thus importantly we have a time-independent, non-Hermitian system exhibiting spontaneously broken $\mathcal{PT}$-symmetry that ordinarily only has a well-defined metric in the unbroken regime. However, we have shown that by introducing time-dependence into this metric, the system becomes well-defined over the \textit{entire} parameter set, including the broken regime. 

The resulting Hermitian Hamiltonian is
\begin{equation}
h\left(t\right)=h_{x,-}\left(t\right)+h_{y,+}\left(t\right),
\end{equation}
where
\begin{equation}
h_{z,\pm}\left(t\right)=\frac{1}{2M_\pm\left(t\right)}p_z^2+\frac{1}{2}M_\pm\left(t\right)\omega_\pm\left(t\right)^2z^2\pm g\left(t\right)\{z,p_z\},
\end{equation}
%\begin{equation}
%h_y\left(t\right)=\frac{1}{2M_y\left(t\right)}p_y^2+\frac{1}{2}M_y\left(t\right)\tilde{\omega}_y\left(t\right)^2y^2-g\left(t\right)\{y,p\}
%\end{equation}
are Swanson type \cite{swanson2004transition} Hamiltonians with time-dependent mass and frequency. The time-dependent terms can be expressed in terms of the Dyson map parameters.

\begin{equation}
M_\pm\left(t\right)=m\left[\cos\theta_++m\alpha_-^2\Gamma_\pm\right]^{-1},
\end{equation}

\begin{equation}
\omega_\pm\left(t\right)^2=\frac{4\Gamma_\mp}{M_\pm},
\end{equation}

\begin{equation}
g\left(t\right)=\frac{\alpha_-\Theta\sin\theta_+}{4},
\end{equation}
where 
\begin{equation}
\Theta\left(t\right)=\frac{2\lambda\sinh\theta_--m\Omega_-^2\cosh\theta_-}{2\cos\theta_++\alpha_+\alpha_-}, \quad \Gamma_\pm\left(t\right)=\frac{1}{16m}\sec\theta_+\left(2m^2\Omega_+^2-\alpha_+^2\right)\pm\frac{\Theta\cos\theta_+}{4}.
\end{equation}
We can recover the time-independent solution by setting $\theta_+\left(t\right)=\alpha_-\left(t\right)=\alpha_+\left(t\right)=0$. In this case $M_\pm\left(t\right)=m$, $\omega_-\left(t\right)=\omega_x$, $\omega_+\left(t\right)=\omega_y$ and $g\left(t\right)=0$. Finally $\theta_-\left(t\right)=\theta$.

As the resulting Hermitian Hamiltonian $h\left(t\right)$ is decoupled, we can solve each system separately following \cite{pedrosa2004complete}. Therefore for $h_{z,\pm}\left(t\right)$ we have
\begin{equation}
\phi_{z,\pm}\left(t\right)=\frac{e^{i\alpha_{n,\pm}\left(t\right)}}{\sqrt{\rho_\pm\left(t\right)}}\exp\left[iM_\pm\left(t\right)\left(\frac{i}{M_\pm\left(t\right)\rho_\pm\left(t\right)^2}+\frac{\dot{\rho}_\pm\left(t\right)}{\rho_\pm\left(t\right)}\mp2g\left(t\right)\right)\frac{z^2}{2}\right]H_n\left[\frac{z}{\rho_\pm\left(t\right)}\right],
\end{equation}
with
\begin{equation}
\alpha_{n,\pm}\left(t\right)=-\left(n+\frac{1}{2}\right)\int^{t}\frac{1}{M_\pm\left(s\right)\rho_\pm\left(s\right)^2}ds,
\end{equation}
and $\rho_\pm$ obeys the dissipative Ermakov-Pinney equation
\begin{equation}
\ddot{\rho}_\pm+\frac{\dot{M}_\pm}{M_\pm}\dot{\rho}_\pm+\left(\omega^2_\pm\mp2\dot{g}-4g^2\mp2g\frac{\dot{M}_\pm}{M_\pm}\right)\rho_\pm=\frac{1}{M^2_\pm\rho^3_\pm}.
\end{equation}
The final solution for $h\left(t\right)$ is therefore
\begin{equation}
\phi\left(t\right)=\phi_{x,-}\left(t\right)\phi_{y,+}\left(t\right).
\end{equation}

\section{Conclusion}

We have calculated a well defined Dyson map $\eta\left(t\right)$ and metric $\varrho\left(t\right)$ for the two dimensional, non-Hermitian, $\mathcal{PT}$-symmetric, coupled oscillator. This Hamiltonian is time-independent and is meaningless without an associated metric. We recalled that a time-independent metric is not valid in the broken regime and then went on to show that a time-dependent metric is valid at the exceptional point, in the broken and unbroken regimes. 

This leads once again to the new interpretation of the placement of time-dependence in quantum mechanics, the metric representation \cite{ExactSols,HigherSpin}, whereby the time-dependence is shifted from the operator to the metric. This is in addition to the Heisenberg and Schr\"odinger representation in which the time dependence is shifted between the operators and the states. The metric representation allows one to calculate observables for systems previously discarded as unphysical and opens up a wide range of possibilities for non-Hermitian Hamiltonians. 

We have demonstrated that the TDDE can be solved for a complex system comprised of ten Hermitian generators. In this case we were not required to take the most general form of $\eta\left(t\right)$ for our ansatz and so there may exist a more general solution if were to explore the entire algebra. \\\\
\textbf{ACKNOWLEDGEMENTS}: TF is supported by a City, University of London Research Fellowship.

%\printbibliography
\bibliography{Bibliography}

\begin{thebibliography}{10}

\bibitem{ExactSols}
A.~Fring and T.~Frith.
\newblock Exact analytical solutions for time-dependent hermitian hamiltonian
  systems from static unobservable non-hermitian hamiltonians.
\newblock \emph{Phys. Rev. A} \textbf{95}(1), 010102 (2017).

\bibitem{fring2016non}
A.~Fring and M.~H. Moussa.
\newblock Non-hermitian swanson model with a time-dependent metric.
\newblock \emph{Phy. Rev. A} \textbf{94}(4), 042128 (2016).

\bibitem{fring2016unitary}
A.~Fring and M.~H. Moussa.
\newblock Unitary quantum evolution for time-dependent quasi-hermitian systems
  with nonobservable hamiltonians.
\newblock \emph{Phys. Rev. A} \textbf{93}(4), 042114 (2016).

\bibitem{FRING20172318}
A.~Fring and T.~Frith.
\newblock Mending the broken $\mathcal{PT}$-regime via an explicit
  time-dependent dyson map.
\newblock \emph{Phys. Lett. A} \textbf{381}(29), 2318 -- 2323 (2017).

\bibitem{HigherDims}
A.~Fring and T.~Frith.
\newblock Solvable two-dimensional time-dependent non-hermitian quantum systems
  with infinite dimensional hilbert space in the broken $\mathcal{PT}$-regime.
\newblock \emph{J. Phys. A: Math. Theor.} \textbf{51}(26), 265301 (2018).

\bibitem{bender1998real}
C.~M. Bender and S.~Boettcher.
\newblock Real spectra in non-hermitian hamiltonians having $\mathcal{PT}$
  symmetry.
\newblock \emph{Phys. Rev. Lett.} \textbf{80}(24), 5243 (1998).

\bibitem{bender2007making}
C.~M. Bender.
\newblock Making sense of non-hermitian hamiltonians.
\newblock \emph{Rep. Prog. Phys.} \textbf{70}(6), 947 (2007).

\bibitem{mostafazadeh2010pseudo}
A.~Mostafazadeh.
\newblock Pseudo-hermitian representation of quantum mechanics.
\newblock \emph{Int. J. Geom. Methods Mod. Phys.} \textbf{7}(7), 1191--1306
  (2010).

\bibitem{moiseyev2011non}
N.~Moiseyev.
\newblock Non-Hermitian quantum mechanics.
\newblock Cambridge University Press (2011).

\bibitem{maamache2017pseudo}
M.~Maamache, O.~K. Djeghiour, N.~Mana, and W.~Koussa.
\newblock Pseudo-invariants theory and real phases for systems with
  non-hermitian time-dependent hamiltonians.
\newblock \emph{Eur. Phys. J} \textbf{132}(9), 383 (2017).

\bibitem{znojil2008time}
M.~Znojil.
\newblock Time-dependent version of crypto-hermitian quantum theory.
\newblock \emph{Phy. Rev. D} \textbf{78}(8), 085003 (2008).

\bibitem{mostafazadeh2007time}
A.~Mostafazadeh.
\newblock Time-dependent pseudo-hermitian hamiltonians defining a unitary
  quantum system and uniqueness of the metric operator.
\newblock \emph{Phys. Lett. B} \textbf{650}(2-3), 208--212 (2007).

\bibitem{de2006time}
C.~F. de~Morisson~Faria and A.~Fring.
\newblock Time evolution of non-hermitian hamiltonian systems.
\newblock \emph{J. Phys. A: Math. Gen.} \textbf{39}(29), 9269 (2006).

\bibitem{mostafazadeh2018energy}
A.~Mostafazadeh.
\newblock Energy observable for a quantum system with a dynamical hilbert space
  and a global geometric extension of quantum theory.
\newblock \emph{arXiv preprint arXiv:1803.04175}  (2018).

\bibitem{PhysRevA.91.062101}
A.~Beygi, S.~P. Klevansky, and C.~M. Bender.
\newblock Coupled oscillator systems having partial $\mathcal{PT}$ symmetry.
\newblock \emph{Phys. Rev. A} \textbf{91}(6), 062101 (2015).

\bibitem{MANDAL20131043}
B.~P. Mandal, B.~K. Mourya, and R.~K. Yadav.
\newblock $\mathcal{PT}$ phase transition in higher-dimensional quantum
  systems.
\newblock \emph{Phys. Lett. A} \textbf{377}(14), 1043 -- 1046 (2013).

\bibitem{scholtz1992quasi}
F.~Scholtz, H.~Geyer, and F.~Hahne.
\newblock Quasi-hermitian operators in quantum mechanics and the variational
  principle.
\newblock \emph{Ann. Phys. (N. Y.)} \textbf{213}(1), 74--101 (1992).

\bibitem{swanson2004transition}
M.~S. Swanson.
\newblock Transition elements for a non-hermitian quadratic hamiltonian.
\newblock \emph{J. Math. Phys.} \textbf{45}(2), 585--601 (2004).

\bibitem{pedrosa2004complete}
I.~Pedrosa.
\newblock Complete exact quantum states of the generalized time-dependent
  harmonic oscillator.
\newblock \emph{Mod. Phys. Lett. B} \textbf{18}(24), 1267--1274 (2004).

\bibitem{HigherSpin}
A.~Fring and T.~Frith.
\newblock Metric versus observable operator representation, higher spin models.
\newblock \emph{Eur. Phys. J} \textbf{133}(2), 57 (2018).

\end{thebibliography}
\bibliographystyle{custom3}

\end{document}